\def\lae{\mathrel{<\kern-1.0em\lower0.9ex\hbox{$\sim$}}}
\def\gae{\mathrel{>\kern-1.0em\lower0.9ex\hbox{$\sim$}}}

\newcommand{\be}{\begin{equation}}
\newcommand{\ee}{\end{equation}}

\documentclass[iop]{emulateapj}
\usepackage{rotating}
\usepackage{amsmath}
\shorttitle{Particle acceleration and synchrotron self-Compton emission in blazar jets \uppercase\expandafter{\romannumeral 1}} \shortauthors{Zheng et al.}

\begin{document}

\title{Particle acceleration and synchrotron self-Compton emission in blazar jets \uppercase\expandafter{\romannumeral 1}: an application to the quiescent emission}
\author{Y. G. Zheng\altaffilmark{1, 2, 3, 4}; S. J. Kang\altaffilmark{5}; C.Y. Yang\altaffilmark{1, 3}; J.M. Bai\altaffilmark{1, 3};}
\altaffiltext{1}{Yunnan Observatories, Chinese Academy of Sciences, Kunming 650011, China (E-mail:baijinming@ynao.ac.cn)}
\altaffiltext{2}{Department of Physics, Yunnan Normal University, Kunming, 650092, China (E-mail:ynzyg@ynu.edu.cn)}
\altaffiltext{3}{Key Laboratory for the Structure and Evolution of Celestial Objects, Chinese Academy of Sciences }
\altaffiltext{4}{Shandong Provincial Key Laboratory of Optical Astronomy and Solar-Terrestrial Environment£¬Shandong University, Weihai, 264209, China}
\altaffiltext{5}{School of Electrical Engineering, Liupanshui Normal University, Liupanshui, Guizhou, 553004, China}

\begin{abstract}
There are still some important unanswered questions about the detailed particle acceleration and escape occurring during the quiescent epoches. As a result, the particle distribution that is adopted in the blazar quiescent spectral model have numerous unconstrained shapes. To help remedy this problem, we introduce a analytical particle transport model to reproduce quiescent broadband spectral energy distribution of blazar. In this model, the exact electron distribution is solved from a generalized transport equation that contains the terms describing first-order and secondary-order \emph{Fermi} acceleration, escape of particle due to both the advection and spatial diffusion, energy losses due to synchrotron emission and inverse-Compton scattering of an assumed soft photon field. We suggest that the advection is a significant escape mechanism in blazar jet. We find that in our model the advection process tends to harden the particle distribution, which enhances the high energy components of resulting synchrotron and synchrotron self-Comptom spectrum from jet. Our model is able to roughly reproduce the observed spectra of extreme BL Lac object 1ES 0414+009 with reasonable assumptions about the physical parameters.
\end{abstract}





\section{Introduction}
Blazars are radio-loud active galactic nucleus (AGN) with a non-thermal continuum emission that arises from the jet emission taking place in an AGN whose jet axis is closely aligned with the observer's line of sight (Ghisellini et al. 1986; Urry \& Padovani 1995). Their broad spectral energy distribution (SED) from the radio to the $\gamma$-rays bands are dominated by two
components, appearing as humps (e.g., Fossati et al. 1998). It is believed that the SED is dominated by various emission mechanisms in different energy regimes (B$\rm \ddot{o}$ttcher 2007). The low-energy hump that extends from radio up to soft X-ray is produced by synchrotron radiation from relativistic electrons and/or positrons in the jet (Urry 1998). Alternatively, in the leptonic model scenarios, the high-energy hump that covers the hard X-ray and $\rm \gamma$-ray energy regime is probably  produced from inverse Compton (IC) scattering of the relativistic electrons either on the synchrotron photons (synchrotron self-Compton, SSC, e.g., Maraschi et al. 1992; Bloom \& Marscher 1996; Mastichiadis \& Kirk 1997; Konopelko et al. 2003) and/or on some other photon populations (external Compton, EC, e.g., Dermer et al. 1992; Dermer \& Schlickeiser. 1993; Sikora et al. 1994; Blandford \& Levinson 1995; Ghisellini \& Madau 1996; B$\rm \ddot{o}$ttcher \& Dermer 1998; Kataoka et al. 1999; Blazejowski et al. 2000; Diltz \& B$\rm \ddot{o}$ttcher 2014; Zheng et al. 2017).

Most of the early models applied to describe quiescent broadband SED of blazars adopt a phenomenological view, assuming that some unspecified mechanism is able to produce the particle distribution that is subsequently injected into the emission region (e.g., Mastichiadis \& Kirk 1997; Bednarek \& Protheroe 1997; 1999; Kataoka et al. 2000; Moderski et al. 2003; Finke et al. 2008; Dermer et al. 2009; Hayashida et al. 2012). The emitting particles required distribution may be established via a variety of mechanisms, including first-order \emph{Fermi} acceleration (shock acceleration) due to multiple shock crossings (e.g., Bell 1978; Blandford \& Ostriker 1978; Drury 1983; Blandford \& Eichler 1987; Jones \& Ellison 1991; Summerlin \& Baring 2012; Marscher 2014; Zheng et al. 2018b), second-order \emph{Fermi} acceleration (stochastic acceleration) due to stochastic interactions with a random field of magnetohydrodynamic (MHD) waves (e.g., Eilek \& Henriksen 1984; Schlickeiser 1984a; 1989; Dung \& Petrosian 1994; Miller \& Roberts 1995; Dermer et al. 1996; Petrosian \& Liu 2004; Katarzynski et al. 2006; Lefa et al. 2011; Zheng \& Zhang 2011; Asano \& Hayashida 2015; Baring et al. 2017), and electrostatic acceleration due to magnetic reconnection (e.g., Giannios et al. 2009; Giannios 2013; Petropoulou et al. 2016; Sironi et al. 2016).

In principle, constructing the particle transport equation and obtaining its solution can produce theoretical SED in the standard blazar paradigm. Previous efforts to solve the particle distribution in some certain assumptions in both an analytical way (e.g., Kardashev 1962; Schlickeiser 1984b, 1985; Park \& Petrosian 1995; Kirk et al. 1998; Keshet \& Waxman 2005; Becker et al. 2006; Stawarz \& Petrosian 2008; Dermer \& Menon 2009; Tramacere et al. 2009; Mertsch 2011; Finke 2013; Lewis et al. 2016; 2018) and a numerical way (Chaiberge \& Ghisellini 1999; Katarzynski et al. 2006; Zheng \& Zhang 2011). However, the behavior of that the particles are trapped in the flow is neglected due to treating escape of particles only as a spatial diffusion. In order to track the advection in the outward direction of jet possible impact on the particle distribution, and then on the photon spectrum, in this paper, we extend the approach introduced by Kroon et al. (2016) from pulsar to the jet of blazars. We focus on a generalized transport equation that contains the terms describing first-order and secondary-order \emph{Fermi} acceleration, escape of particle due to both the advection and spatial diffusion, energy losses due to synchrotron emission and IC scattering of an assumed soft photon field. Our main aim is to show that the particle distribution in the context is able to reproduce the multi-wavelength spectrum with reasonable assumptions about the physical parameters.

The present paper is organized as follows. In Section 2 we describe the transport equation that contains the terms of first-order \emph{Fermi} acceleration, secondary-order \emph{Fermi} acceleration, particle escape, and energy losses. In Section 3 we compare the timescales of first-order \emph{Fermi} acceleration, secondary-order \emph{Fermi} acceleration, particle escape, and energy losses. In Section 4 we solve the steady-state transport equation to obtain the solution of particle Green's function. In Section 5 we deduce the particle Green's function in a special case of low particle momentum. In Section 6 we calculate the theoretical photon spectrum utilizing the particle Green's function. In Section 7 we apply the model to the quiescent state emission from extreme BL Lac object 1ES 0414+009, and some discussions are be given in Section 8. Throughout the paper, we assume the Hubble constant $H_{0}=75$ km s$^{-1}$ Mpc$^{-1}$, the dimensionless numbers for the energy density of matter $\Omega_{\rm M}=0.27$, the dimensionless numbers of radiation energy density $\Omega_{\rm r}=0$, and the dimensionless cosmological constant $\Omega_{\Lambda}=0.73$.

\section{Basic equations}
Assuming the energetic particles in a turbulent and tenuous plasma carrying a magnetic field, the momentum spectrum of particles undergoing \emph{Fermi} acceleration due to irregularly moving magnetized fluid elements can be studied in terms of a diffusion equation in momentum space (Tverskoi 1967; Tsytovich 1977)

\begin{equation}
\frac{\partial f(p,t)}{\partial t}=\frac{1}{p^{2}}\frac{\partial}{\partial p}\biggl[p^{2}D(p)\frac{\partial f(p,t)}{\partial p}\biggr]+Q(p,t),
\label{Eq:1}
\end{equation}
where $f(p,t)$ is the isotropic, homogeneous phase space density, $p$ the particle momentum, and $D(p)$ the second order \emph{Fermi} acceleration diffusion coefficient by scattering off MHD waves, $Q(p,t)$ the sources and sinks of particles. The phase space density is related to the total number of particles, $N_{e}$, via $N_{e}(t)=\int^{\infty}_{0}4\pi p^{2}f(p,t)dp$.

It is well known that the formation of strong shock can be expected in the jet of blazar around locations a few parsecs from the core (Edwards \& Piner 2002; Piner et al. 2009). The particles can gain momentum by first order \emph{Fermi} acceleration off the strong shocks (Bell 1978; Axford 1981). While the particle gain momentum from both the shock and turbulence, they also suffer from many kinds of momentum loss processes. Incorporating these effects with escape into Eq. {\ref{Eq:1}}, we obtain

\begin{eqnarray}
\frac{\partial f(p,t)}{\partial t}&=&\frac{1}{p^{2}}\frac{\partial}{\partial p}\biggl\{p^{2}\biggl[D(p)\frac{\partial f(p,t)}{\partial p}-\dot{p}_{\rm gain}f(p,t)\nonumber\\&-&\dot{p}_{\rm loss}f(p,t)\biggr]\biggr\}-\frac{f(p,t)}{t_{\rm esc}(p)}+Q(p,t)\,.
\label{Eq:2}
\end{eqnarray}

\subsection{Stochastic Acceleration}
The turbulent magnetic field component gives rise to spatial diffusion of charged plasma particles which is described by the spatial diffusion coefficient $\kappa(p)$. In the case of an isotropic Alfvenic turbulence with one dimensional power spectrum $W(k)\propto k^{-q}$ with a spectral index $q$ in a finite wave-vector range $k_{\rm min}<k<k_{\rm max}$, the relation between the spatial diffusion coefficient and the momentum diffusion coefficient can be written as (Skilling 1975; Webb 1983; Schlickeiser 1985; Dr$\rm \ddot{o}$ge et al. 1987)
\begin{equation}
D(p)=\frac{v_{A}^{2}p^{2}}{9\kappa(p)}\,,
\label{Eq:3}
\end{equation}
where $v_{A}$ is the Alfven velocity. In the case of particles with larmor radius, $r_{\rm L}$, smaller than the correlation length of the field, we can introduce a dimensionless parameter
\begin{equation}
\eta=(\frac{B}{\delta B})^{2}(\frac{\lambda_{\rm max}}{r_{\rm L}})^{(q-1)}\,,
\label{Eq:4}
\end{equation}
where $B$ is the local magnetic field strength, $\delta B$ the turbulent component of the magnetic field, $\lambda_{\rm max}=2\pi/k_{\rm min}$ the maximum wavelength of the Alfven modes, to parameterize the particle mean-free path, $\ell$, relative to the larmor radius by
\begin{equation}
\ell=\eta r_{\rm L}=\eta\frac{pc}{eB}\,,
\label{Eq:5}
\end{equation}
where $c$ is the speed of light, and $e$ the magnitude of the electron charge.

The associated spatial diffusion coefficient with mean-free path is calculated using (Reif 1965; Dr$\rm \ddot{o}$ge \& Schlickeiser 1986)
\begin{equation}
\kappa(p)=\frac{c\ell}{3}\,.
\label{Eq:6}
\end{equation}
Combining Eq.({\ref{Eq:3}}), Eq.({\ref{Eq:5}}), and Eq.({\ref{Eq:6}}), we find that the momentum diffusion coefficient can be given as (e.g. Dermer et al. 1996; Becker et al. 2006)
\begin{equation}
D(p)=D_{0}m_{e}cp\,,
\label{Eq:7}
\end{equation}
with a momentum diffusion rate constant
\begin{equation}
D_{0}=\frac{eB\sigma_{\rm mag}}{3\eta m_{e}c}=5.86\times10^{5}\sigma_{\rm mag}\eta^{-1}\biggl(\frac{B}{0.1~\rm G}\biggr)~~\rm s^{-1}\,,
\label{Eq:8}
\end{equation}
where $\sigma_{\rm mag}=v_{A}^{2}/c^{2}$ is the magnetization parameter (e.g. Sironi \& Spitkovsky 2014). This relation gives the stochastic momentum gain rate as (Becker et al. 2006)
\begin{equation}
\dot{p}_{\rm stoch}=\frac{1}{p^{2}}\frac{\partial}{\partial p}[p^{2}D(p)]=3D_{0}m_{e}c\,.
\label{Eq:9}
\end{equation}
\subsection{Shock Acceleration}
We consider quasi-continuous momentum gain by systematic acceleration at shock waves moving through the plasma at speed $v_{s}$. The momentum gain rate by shock acceleration at an isolated single shock waves is determined (e.g. Drury 1983; Lagage \& Cesarsky 1983)
\begin{equation}
\dot{p}_{\rm sh}=\frac{1}{3}(U_{1}-U_{2})\biggl[\frac{\kappa_{1}}{U_{1}}+\frac{\kappa_{2}}{U_{2}}\biggr]^{-1}p\simeq\frac{v_{s}^{2}}{4\kappa(p)}p\,,
\label{Eq:10}
\end{equation}
where $U_{1}~(U_{2})$ and $\kappa_{1}~(\kappa_{2})$ are the flow velocities and diffusion coefficients upstream (downstream) of the chock in the shock's comoving frame. In order to simplify the model, we attribute the momentum gain rate in Eq. (2) only to the shock acceleration. Combining Eq.({\ref{Eq:5}}), Eq.({\ref{Eq:6}}), and Eq.({\ref{Eq:10}}), we find that the momentum gain rate experienced by the particles due to multiple shock crossings can be given as
\begin{equation}
\dot{p}_{\rm gain}=\dot{p}_{\rm sh}=A_{0}m_{e}c\,,
\label{Eq:11}
\end{equation}
with a shock acceleration rate constant
\begin{equation}
A_{0}=\frac{3\xi eB}{4m_{e}c}=1.32\times10^{6}\xi\biggl(\frac{B}{0.1~\rm G}\biggr)~~\rm s^{-1}\,,
\label{Eq:12}
\end{equation}
where $\xi=\eta^{-1}v_{s}^{2}/c^{2}$ is an efficiency factor.
\subsection{Momentum loss}
In the presence of ambient magnetic, photon fields in the dissipated region of jet, the particles also undergo synchrotron radiation, inverse Compton scattering (ICs). The synchrotron and ICs energy loss rate per particle, averaged over an isotropic distribution of pitch angles, are given by (e.g. Rybicki \& Lightman)
\begin{equation}
\dot{\gamma}_{\rm syn}m_{e}c^{2}=-\frac{4}{3}\sigma_{T}cu_{B}\gamma^{2}\,,
\label{Eq:13}
\end{equation}
and
\begin{equation}
\dot{\gamma}_{\rm ICs}m_{e}c^{2}=-\frac{4}{3}\sigma_{T}cu_{\rm ph}\gamma^{2}\,,
\label{Eq:14}
\end{equation}
respectively. Here $\sigma_{T}$ is the Thomson cross section, $u_{B}=B^{2}/8\pi$ the magnetic field density, and $u_{ph}$ the soft photon density to be up-scattered. The associated synchrotron and ICs momentum loss rate can be written as
\begin{equation}
\dot{p}_{\rm syn}=-\frac{4\sigma_{T}u_{B}}{3m_{e}c}\frac{p^{2}}{m_{e}c}\,,
\label{Eq:15}
\end{equation}
and
\begin{equation}
\dot{p}_{\rm ICs}=-\frac{4\sigma_{T}u_{\rm ph}}{3m_{e}c}\frac{p^{2}}{m_{e}c}\,.
\label{Eq:16}
\end{equation}
Hence the momentum loss rate, $\dot{p}_{\rm loss}$, appearing in the Eq. (\ref{Eq:2}) can be written as the sum
\begin{equation}
\dot{p}_{\rm loss}=\dot{p}_{\rm syn}+\dot{p}_{\rm ICs}=-\frac{B_{0}}{m_{e}c}p^{2}\,,
\label{Eq:17}
\end{equation}
where, the momentum loss rate constant $B_{0}$ is given by
\begin{equation}
B_{0}=\frac{4\sigma_{T}}{3m_{e}c}(u_{B}+u_{\rm ph})=3.25\times10^{-8}u~~\rm s^{-1}\,,
\label{Eq:18}
\end{equation}
with a constant soft photon field $u_{\rm ph}$.
\subsection{Escape of Particles}
The escape in the calculations can generally be explained in a real process of the particles into the region of a source where the magnetic field strength is significantly smaller and therefore the efficiency of the particle emission is also significantly less (e.g. Katarzynski et al. 2006). In this scenario, the particles remain in the acceleration region for a mean times $t_{\rm esc}$ before escaping. In order to properly treat as the dominant spatial transport processes on large and small scales, we introduce a momentum dependence escape time-scale, $t_{\rm esc}(p)$.

It is believe that the Larmor radius of a particle with small momenta is much smaller than the size of acceleration region. In this case, the particle is trapped in the flow, and the escape of particles from the acceleration region occurs via advection (Becker \& Begelman 1986). This process is called shock regulated escape (Steinacker \& Schlickeiser 1989) with a timescale, $t_{\rm SRE}(p)$, as (e.g. Jokipii 1987; Gallant \& Achterberg 1999)
\begin{equation}
t_{\rm SRE}(p)=\frac{p}{C_{0}m_{e}c}\,,
\label{Eq:19}
\end{equation}
where, $C_{0}$ is the shock regulated escape rate constant with
\begin{equation}
C_{0}=\frac{eB}{\omega m_{e}c}=1.76\times10^{6}\omega^{-1}\biggl(\frac{B}{0.1~\rm G}\biggr)~~\rm s^{-1}\,.
\label{Eq:20}
\end{equation}
Here, $\omega$ is a dimensionless constant of order of unity that accounts for time dilation and obliquity in the relativistic shock (Kroon et al. 2016).

On the contrary, in the case of a particle with large momenta, the escape of the particles occurs via spatial diffusion. This process is called \emph{Bohm} diffusive escape (Dermer \& Menon 2009) with a timescale (Kroon et al. 2016),
\begin{equation}
t_{\rm Bohm}(p)=\frac{m_{e}c}{F_{0}p}\,,
\label{Eq:21}
\end{equation}
where, $F_{0}$ is the \emph{Bohm} diffusive escape rate constant with
\begin{eqnarray}
F_{0}&=&\frac{\eta m_{e}c^{3}}{r_{s}^{2}eB}\nonumber\\&=&5.12\times10^{-20}\eta\biggl(\frac{r_{s}}{10^{17}~\rm cm}\biggr)^{-2}\biggl(\frac{B}{0.1~\rm G}\biggr)^{-1}~~\rm s^{-1}\,.
\label{Eq:22}
\end{eqnarray}
Here, $r_{s}$ is the size of blob.

It can be seen that either the particles with small momentum are likely to advect away into the downstream region, or the particles with large momentum are likely to diffuse out of the acceleration region via \emph{Bohm} diffusion. In order to ensure that the behavior of particles with both the small and large momentum are properly taken into account, these two escape rates can be included in the net escape rate, $t^{-1}_{\rm esc}(p)$, given by
\begin{equation}
t^{-1}_{\rm esc}(p)=t^{-1}_{\rm SRE}(p)+t^{-1}_{\rm Bohm}(p)\,.
\label{Eq:23}
\end{equation}
Inserting Eq.(\ref{Eq:19}) and Eq.(\ref{Eq:21}) into Eq.(\ref{Eq:23}), we find
\begin{equation}
t_{\rm esc}(p)=\biggl(\frac{C_{0}m_{e}c}{p}+\frac{F_{0}p}{m_{e}c}\biggr)^{-1}\,.
\label{Eq:24}
\end{equation}

\subsection{Particles Injection}
The injection process create a seed population of the non-thermal particle, while the particle injection mechanism is not well understood. Since the model considered here includes significant components of particle acceleration and cooling, the evolution of momentum distribution is independent of the precise form of the momentum distribution of the injected electrons (e.g. Katarzynski et al. 2006; Zheng \& Zhang 2011). We can utilize this insensitivity by assuming that the injected particles have a mono-energetic distribution, with a characteristic injection momentum, $p_{0}$,
\begin{equation}
Q(p,t)=\frac{\dot{N}_{0}\delta(p-p_{0})}{4\pi p_{0}^{2}}\,,
\label{Eq:25}
\end{equation}
where $\dot{N}_{0}$ is the continual injection rate in the units of $~p^{-1}\rm cm^{-3}~s^{-1}$, and $\delta(p)$ the Dirac's distribution function.

\subsection{Transport Equation}
Substituting Eq. (\ref{Eq:7}), Eq. (\ref{Eq:11}), Eq. (\ref{Eq:17}), Eq. (\ref{Eq:24}) and Eq. (\ref{Eq:25}) into Eq. (\ref{Eq:2}), we find the basic transport equation given by
\begin{eqnarray}
\frac{\partial f(p,t)}{\partial t}&=&\frac{1}{p^{2}}\frac{\partial}{\partial p}\biggl\{p^{2}\biggl[D_{0}m_{e}cp\frac{\partial f(p,t)}{\partial p}-A_{0}m_{e}c f(p,t)\nonumber\\&+&\frac{B_{0}p^{2}}{m_{e}c}f(p,t)\biggr]\biggr\}-\biggl(\frac{C_{0}m_{e}c}{p}+\frac{F_{0}p}{m_{e}c}\biggr)f(p,t)\nonumber\\&+&\frac{\dot{N}_{0}\delta(p-p_{0})}{4\pi p_{0}^{2}}\,.
\label{Eq:26}
\end{eqnarray}
This equation can be used to calculate the particle distribution in emission region of blazar jet by both analytical (e.g. Schlickeiser 1984b, 1985; Park \& Petrosian 1995; Becker et al. 2006; Stawarz \& Petrosian 2008; Mertsch 2011; Zheng et al. 2018a) and numerical approaches (e.g. Chaiberge \& Ghisellini 1999; Katarzynski et al. 2006; Zheng \& Zhang 2011).
\section{The Timescales}
The timescale determines how long it takes for a particle to gain or loss momentum, allowing easy comparison between the efficiencies of different gain and loss mechanisms. When the gain or loss rate of momentum is determined, we can estimate the timescale by
\begin{equation}
t(p)=\frac{p}{\dot{p}}\,.
\label{Eq:27}
\end{equation}

\begin{figure}
	\centering
		\includegraphics[width=9.0 cm]{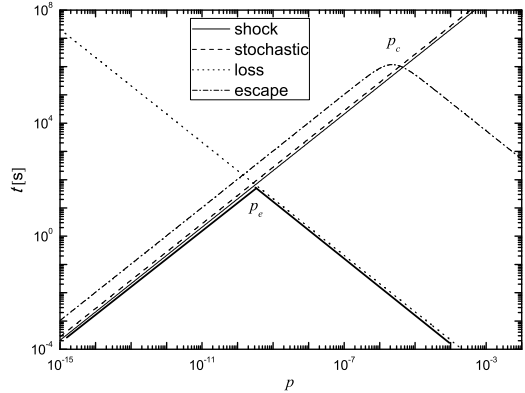}
		
	\caption{Calculated timescales in the comoving frame of the plasma as a function of the particle momentum. The solid curve represent shock acceleration, the dashed curve represent stochastic acceleration, the dotted curve represent momentum loss due to synchrotron emission and ICs, and the dash-dotted curve represent the escape of particles. The thick black curve show the dominant process. The $p_{e}$ is the equilibrium momentum and $p_{c}$ is the cross-over momentum. We adopt the parameters as follows: $B=0.1~\rm G$, $u=0.398~\rm erg~cm^{-3}$, $\sigma_{\rm mag}=0.1$, $\eta=1.0$, $\xi=0.1$, $\omega=0.1$, and $r_{s}=1.0\times10^{16}~\rm cm$.}
	\label{fig:1}
\end{figure}

We show the calculated timescales in the comoving frame of the plasma as a function of the particle momentum in Figure {\ref{fig:1}}. It is believed that the effective escape timescale is $t_{\rm esc}\rightarrow0$ in the limits of $p\rightarrow0$ and $p\rightarrow\infty$. In this scenario, we can find the cross-over momentum with $t_{\rm SRE}(p)=t_{\rm Bohm}(p)$ as
\begin{eqnarray}
p_{c}&=&m_{e}c\sqrt{\frac{C_{0}}{F_{0}}}\nonumber\\&=&1.6\times10^{-4}\eta^{-1}\omega^{-1}\biggl(\frac{r_{s}}{10^{17}~\rm cm}\biggr)\biggl(\frac{B}{0.1~\rm G}\biggr)\,.
\label{Eq:28}
\end{eqnarray}
It can be seen that, in the regime of $p<p_{c}$, the escape of particles is dominated by advection, and in the regime of $p>p_{c}$, the escape of particles is dominated by spatial diffusion. The cross-over momentum can determine a critical Lorentz factor by $\gamma_{c}=p_{c}/(m_{e}c)$. If we adopt a typical values of $r_{s}\sim10^{16}~\rm cm$ and $B\sim0.1~\rm G$ for a blazar jet, we find $\gamma_{c}\sim10^{12}$. This is far from the Lorentz factor, which yields TeV $\gamma$-ray photons by SSC processes in the jet. As an open issue, we suggest that the advection is a significant escape mechanism in blazar jet. This issue tends to harden the particle distribution, which enhances the high energy components of resulting synchrotron and SSC spectrum from jet.

On the other hand, a dynamic equilibrium is generated by a kind of competition between the acceleration, injection, escape and the cooling of particles from the shock region. Since the escape of particles is dominated by advection in blazar jet, we can expect a theoretical equilibrium momentum, $p_{e}$, by $t_{\rm loss}(p)=min[t_{\rm stoch}(p), t_{\rm gain}(p), t_{\rm SRE}(p)]$. Relativistic particles with a equilibrium momentum may be responsible for the X-ray and TeV $\gamma$-ray photons in the SSC framework. These can provide a rough estimate for the time it takes the electron distribution to reach equilibrium.

\section{A Stationary Particles Distribution}
In despite of the variability, which is found from radio to TeV $\gamma$-ray bands, is one of the major characteristics of blazars, these sources should persists the quiescent state throughout many epoches. In this scenario, we expect a stationary particle distribution and radiation in the emission region. In order to do so, we set $\partial f(p,t)/\partial t=0$ in Eq.(\ref{Eq:26}), and then solve a stationary particle transport equation.

We first solve the steady-state Green's function, $f_{G}(p, p_{0})$, with a given source distribution $Q(p)$ under the proper boundary condition (e.g. Schlickeiser 1984b). Once the Green's function is determine, the steady-state density $f(p)$ can be obtained using the convolution
\begin{equation}
f(p)=\int_{0}^{\infty}f_{G}(p,p_{0})\delta(p-p_{0})d p_{0}\,.
\label{Eq:29}
\end{equation}
We define the dimensionless momentum, $\upsilon=p/(m_{e}c)$, and the dimensionless time, $\tau=D_{0}t$. Reminding the characteristics of Dirac's function, we combine Eq. (\ref{Eq:26}) and Eq. (\ref{Eq:29}). In case of the new coordinates ($\upsilon, \tau$), the steady-state transport equation can be written as (e.g. Kroon et al. 2016)

\begin{eqnarray}
\frac{\partial f_{G}(\upsilon,\upsilon_{0})}{\partial\tau}&=&\frac{1}{\upsilon^{2}}\frac{\partial}{\partial\upsilon}\biggl\{\upsilon^{2}\biggl[\upsilon\frac{\partial f_{G}(\upsilon,\upsilon_{0})}{\partial\upsilon}-\hat{A}f_{G}(\upsilon,\upsilon_{0})\nonumber\\&+&\hat{B}\upsilon^{2}f_{G}(\upsilon,\upsilon_{0})\biggr]\biggr\}-
\biggl(\frac{\hat{C}}{\upsilon}+\hat{F}\upsilon\biggr)f_{G}(\upsilon,\upsilon_{0})\nonumber\\&+&\frac{\dot{N}_{0}m_{e}c\delta(\upsilon-\upsilon_{0})}{4\pi D_{0}\upsilon_{0}^{2}}=0\,,
\label{Eq:30}
\end{eqnarray}
where, we let $\hat{A}=A_{0}/D_{0}$, $\hat{B}=B_{0}/D_{0}$, $\hat{C}=C_{0}/D_{0}$, and $\hat{F}=F_{0}/D_{0}$.

It is convenient to relate the phase space density, $f_{G}(\upsilon, \upsilon_{0}, \tau)$, to particle number density, $N_{G}(\upsilon, \upsilon_{0}, \tau)$, using
\begin{equation}
f_{G}(\upsilon, \upsilon_{0}, \tau)=\frac{N_{G}(\upsilon, \upsilon_{0}, \tau)}{4\pi\upsilon^{2}}\,.
\label{Eq:31}
\end{equation}
Substituting Eq. (\ref{Eq:31}) into Eq. (\ref{Eq:30}), we obtain
\begin{eqnarray}
\frac{\partial N_{G}(\upsilon,\upsilon_{0})}{\partial\tau}&=&\frac{\partial}{\partial\upsilon}\biggl[N_{G}(\upsilon,\upsilon_{0})+\upsilon\frac{\partial N_{G}(\upsilon,\upsilon_{0})}{\partial\upsilon}\biggr]\nonumber\\&+&\frac{\partial}{\partial\upsilon}\biggl[(\hat{B}\upsilon^{2}-3-\hat{A})N_{G}(\upsilon,\upsilon_{0})\biggr]\nonumber\\
&-&\biggl(\frac{\hat{C}}{\upsilon}+\hat{F}\upsilon\biggr)N_{G}(\upsilon,\upsilon_{0})\nonumber\\&+&\frac{\dot{N}_{0}m_{e}c\delta(\upsilon-\upsilon_{0})}{D_{0}}=0\,.
\label{Eq:32}
\end{eqnarray}
Since $N_{G}(\upsilon,\upsilon_{0})$ is independent from the $\tau$, we can rewrite Eq. (\ref{Eq:32}) as
\begin{eqnarray}
&\upsilon&\frac{d^{2}N_{G}(\upsilon,\upsilon_{0})}{d\upsilon^{2}}+(\hat{B}\upsilon^{2}-1-\hat{A})\frac{d N_{G}(\upsilon,\upsilon_{0})}{d\upsilon}\nonumber\\
&+&(2\hat{B}\upsilon-\frac{\hat{C}}{\upsilon}-\hat{F}\upsilon)N_{G}(\upsilon,\upsilon_{0})=-\frac{\dot{N}_{0}m_{e}c\delta(\upsilon-\upsilon_{0})}{D_{0}}\,.
\label{Eq:33}
\end{eqnarray}
The Green's function, $N_{G}(\upsilon,\upsilon_{0})$, must be continuous at the momentum $\upsilon=\upsilon_{0}$. The derivative shows a jump that can be deduced by integrating Eq. (\ref{Eq:33}) with respect to $\upsilon$ over a small region $\delta_{0}$ around $\upsilon_{0}$ (e.g. Kroon et al. 2016). The integration results to
\begin{equation}
\underset{\delta\rightarrow0}{\lim} \frac{d N_{G}(\upsilon, \upsilon_{0})}{d\upsilon}\biggr|_{\upsilon_{0}+\delta_{0}}-\underset{\delta\rightarrow0}{\lim} \frac{d N_{G}(\upsilon, \upsilon_{0})}{d\upsilon}\biggr|_{\upsilon_{0}-\delta_{0}}=-\frac{\dot{N}_{0}m_{e}c}{D_{0}\upsilon_{0}}\,.
\label{Eq:34}
\end{equation}
At the the momentum $\upsilon\neq\upsilon_{0}$, Eq. (\ref{Eq:33}) can be rewritten as a confluent hypergeometric function (Kurmmer 1837)
\begin{eqnarray}
\upsilon\frac{d^{2}N_{G}(\upsilon,\upsilon_{0})}{d\upsilon^{2}}&+&(\hat{B}\upsilon^{2}-1-\hat{A})\frac{d N_{G}(\upsilon,\upsilon_{0})}{d\upsilon}\nonumber\\
&+&(2\hat{B}\upsilon-\frac{\hat{C}}{\upsilon}-\hat{F}\upsilon)N_{G}(\upsilon,\upsilon_{0})=0\,.
\label{Eq:35}
\end{eqnarray}
In this case, the Green's function, $N_{G}(\upsilon,\upsilon_{0})$, fulfils appropriate boundary conditions at both $\upsilon\rightarrow0$ and $\upsilon\rightarrow\infty$ (Tademaru et al. 1971; Melrose 1971; Bicknell \& Melrose 1982). It can be expressed in terms of the Whittaker function $M_{\sigma, \mu}$ and $W_{\sigma, \mu}$ (Abramowitz \& Stegun 1970)
\begin{equation}
N_{G}(\upsilon, \upsilon_{0})\propto \upsilon^{\frac{\hat{A}}{2}}e^{-\frac{\hat{B}\upsilon^{2}}{4}}
\begin{cases}
M_{\sigma, \mu}(\frac{\hat{B}\upsilon^{2}}{2}), &\text{$\upsilon\leq\upsilon_{0}$},\\
W_{\sigma, \mu}(\frac{\hat{B}\upsilon^{2}}{2}), &\text{$\upsilon>\upsilon_{0}$},
\end{cases}
\label{Eq:36}
\end{equation}
where, we define the parameters $\sigma=1+\hat{A}/4-\hat{F}/(2\hat{B})$, and $\mu=0.25[(2+\hat{A})^{2}+4\hat{C}]^{1/2}$. Taking into account the continuity of the Green's function at the momentum $\upsilon=\upsilon_{0}$, Kroon et al. (2016) gives the particle Green's function
\begin{eqnarray}
N_{G}(\upsilon, \upsilon_{0})&=&\frac{\dot{N}_{0}m_{e}c\Gamma(\mu-\sigma+0.5)}{\hat{B}D_{0}\Gamma(1+2\mu)\upsilon_{0}^{2}}\biggl(\frac{\upsilon}{\upsilon_{0}}\biggr)^{\frac{\hat{A}}{2}} e^{-\frac{\hat{B}(\upsilon^{2}-\upsilon_{0}^{2})}{4}}\nonumber\\&\times&M_{\sigma, \mu}(\frac{\hat{B}\upsilon_{\rm 1}^{2}}{2})W_{\sigma, \mu}(\frac{\hat{B}\upsilon_{\rm 2}^{2}}{2})\,,
\label{Eq:37}
\end{eqnarray}
with $\upsilon_{\rm 1}=\rm min[\upsilon, \upsilon_{0}]$, and $\upsilon_{\rm 2}=\rm max[\upsilon, \upsilon_{0}]$, where $\Gamma(x)$ is the Gamma's function. This equation exhibits the particle distribution resulting from a dynamic equilibrium between the acceleration, injection, escape and the cooling of particles.

We can utilize the relation, $E_{e}^{2}=p^{2}c^{2}+m_{e}^{2}c^{2}$, to connect the non-thermal particle energy $E_{e}$ and the momentum in general. This gives the relationship $\upsilon=\sqrt{\gamma^{2}-1}$ between $\upsilon$ and the Lorentz factor of particles $\gamma$. Since the ultra-relativistic particles ($\gamma\gg1$) dominates on the SEDs of blazars, we can write $\upsilon=\gamma$ without making significant error. In this scenario, we can rewrite Eq. (\ref{Eq:37}) as
\begin{eqnarray}
N_{G}(\gamma, \gamma_{0})&=&\frac{\dot{N}_{0}m_{e}c\Gamma(\mu-\sigma+0.5)}{\hat{B}D_{0}\Gamma(1+2\mu)\gamma_{0}^{2}}\biggl(\frac{\gamma}{\gamma_{0}}\biggr)^{\frac{\hat{A}}{2}} e^{-\frac{\hat{B}(\gamma^{2}-\gamma_{0}^{2})}{4}}\nonumber\\&\times&M_{\sigma, \mu}(\frac{\hat{B}\gamma_{\rm 1}^{2}}{2})W_{\sigma, \mu}(\frac{\hat{B}\gamma_{\rm 2}^{2}}{2})\,.
\label{Eq:38}
\end{eqnarray}

\begin{figure}
	\centering
		\includegraphics[width=9.0 cm]{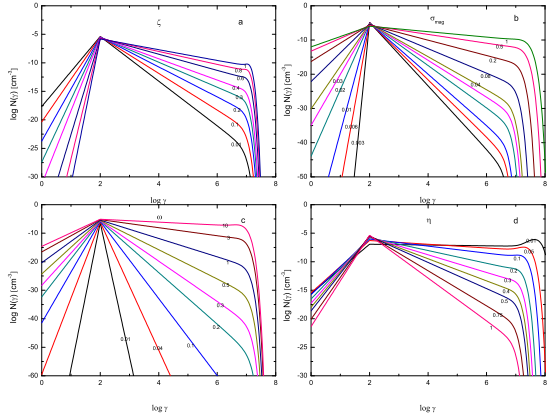}
		
	\caption{Example of particle distribution for four dimensionless parameter changes. The top panels show the shock acceleration efficiency factor $\xi$ (a) and the magnetization parameter $\sigma_{\rm mag}$ (b). The bottom panels show the time dilation and obliquity in the relativistic shock $\omega$ (c) and the parameter of connecting particle mean-free path to Larmor radius $\eta$ (d). Marks near color curves represent the values of the parameter. We adopt the other parameters as follows: $\dot{N}_{0}=3.0~p^{-1}~\rm cm^{-3}~s^{-1}$, $B=0.1~\rm G$, $u=3.98~\rm erg~cm^{-3}$, $\gamma_{0}=100$, and $r_{s}=1.0\times10^{16}~\rm cm$.}
	\label{fig:2}
\end{figure}

The particle distribution given by Eq. (\ref{Eq:38}) can be used to calculate the theoretical SED produced from a population of radiating relativistic particles in the blazar jet under the combined action of stochastic acceleration, shock acceleration, particle escape, synchrotron and ICs losses. We show the particle distribution with different parameters in Figure \ref{fig:2}. It can be seen that: 1) in the regime of $\gamma\leq\gamma_{e}=p_{e}/(m_{e}c)$, the particle distributions show a cusp centered at the injection Lorentz factor, $\gamma_{0}=p_{0}/(m_{e}c)$, surrounded by two power-law wings with different spectral index. The indices of power-law wings are sensitive to the four dimensionless parameters; 2) in the regime of $\gamma>\gamma_{e}$, where the energy losses overwhelm the particle acceleration, the particle distribution terminates in an exponential cutoff.

\section{A Special Case of Lower Particle Momentum}
It is interesting to note that, when the particle momentum satisfies $p\ll p_{e}$, the processes of energy loss and \emph{Bohm} diffusion escape should be neglected. In this special case, we find that the Eq. (\ref{Eq:33}) reduces to an \emph{Euler} equidimensional equation as
\begin{eqnarray}
\upsilon^{2}\frac{d^{2}N_{G}(\upsilon,\upsilon_{0})}{d\upsilon^{2}}&-&(1+\hat{A})\upsilon\frac{d N_{G}(\upsilon,\upsilon_{0})}{d\upsilon}
\nonumber\\&-&\hat{C}N_{G}(\upsilon,\upsilon_{0})=-\frac{\dot{N}_{0}\upsilon\delta(\upsilon-\upsilon_{0})}{D_{0}}\,.
\label{Eq:39}
\end{eqnarray}
At the the momentum $\upsilon\neq\upsilon_{0}$, Eq. (\ref{Eq:39}) can be rewritten as a homogeneous \emph{Euler} equidimensional equation of the form
\begin{eqnarray}
\upsilon^{2}\frac{d^{2}N_{G}(\upsilon,\upsilon_{0})}{d\upsilon^{2}}&-&(1+\hat{A})\upsilon\frac{d N_{G}(\upsilon,\upsilon_{0})}{d\upsilon}
\nonumber\\&-&\hat{C}N_{G}(\upsilon,\upsilon_{0})=0\,.
\label{Eq:40}
\end{eqnarray}
Using the change of variables with $\upsilon=e^{\iota}$, we can obtain a power-law solutions of the form
\begin{equation}
N_{G}(\upsilon, \upsilon_{0})=H_{0}\upsilon^{\alpha},
\label{Eq:41}
\end{equation}
for Eq. (\ref{Eq:40}) (e.g. Alzate et al. 2016), where $H_{0}$ is a normalization constant and $\alpha$ an power-law index. We can determine the power-law index using the characteristic polynomial
\begin{equation}
\alpha^{2}-(2+\hat{A})\alpha-\hat{C}=0,
\label{Eq:42}
\end{equation}
where the both roots
\begin{equation}
\alpha_{1}=\frac{2+\hat{A}+\sqrt{(2+\hat{A})^{2}+4\hat{C}}}{2}\,,
\label{Eq:43}
\end{equation}
applies in the low momentum regime with $\upsilon\le\upsilon_{0}$, and
\begin{equation}
\alpha_{2}=\frac{2+\hat{A}-\sqrt{(2+\hat{A})^{2}+4\hat{C}}}{2}\,,
\label{Eq:44}
\end{equation}
applies in the high momentum regime with $\upsilon>\upsilon_{0}$. Application of the derivative jump condition given by Eq. (\ref{Eq:34}),  Kroon et al. (2016) gives the properly normalized global solution
\begin{eqnarray}
N_{G}(\upsilon, \upsilon_{0})&=&\frac{\dot{N}_{0}m_{e}c}{4D_{0}\mu}
\begin{cases}
(\frac{\upsilon}{\upsilon_{0}})^{\alpha_{1}}, &\text{$\upsilon\leq\upsilon_{0}$},\\
(\frac{\upsilon}{\upsilon_{0}})^{\alpha_{2}}, &\text{$\upsilon>\upsilon_{0}$}.
\end{cases}
\label{Eq:45}
\end{eqnarray}
Reminding the relation between momentum and energy, we can rewrite Eq. (\ref{Eq:45}) as
\begin{eqnarray}
N_{G}(\gamma, \gamma_{0})&=&\frac{\dot{N}_{0}m_{e}c}{4D_{0}\mu}
\begin{cases}
(\frac{\gamma}{\gamma_{0}})^{\alpha_{1}}, &\text{$\gamma\leq\gamma_{0}$},\\
(\frac{\gamma}{\gamma_{0}})^{\alpha_{2}}, &\text{$\gamma>\gamma_{0}$}.
\end{cases}
\label{Eq:46}
\end{eqnarray}

In Figure 3, we compare the particle distribution with energy loss and \emph{Bohm} diffusion escape, exhibiting in Eq. (\ref{Eq:38}), and without energy loss and \emph{Bohm} diffusion escape, exhibiting in Eq. (\ref{Eq:46}). As shown in Figure 3, the effect of energy loss is to move high-energy particles to lower energies, resulting in an increased curvature and a steepened particle distribution at high energy regimes. Since both the shock regulated escape and acceleration processes tend to harden the particle spectrum, we expect a powerful high-energy components of resulting synchrotron and SSC spectrum from jet.
\begin{figure}
	\centering
		\includegraphics[width=9.0 cm]{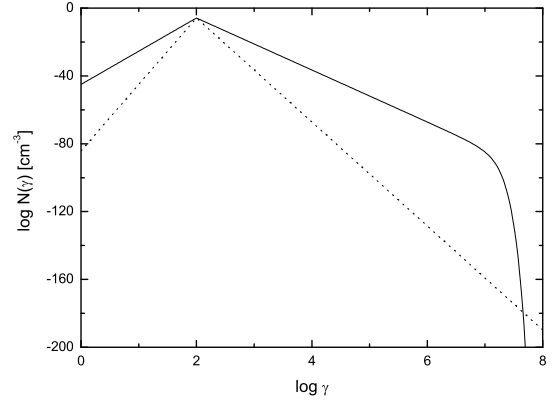}
		
	\caption{Comparison between the particle distribution with energy loss and \emph{Bohm} diffusion escape (solid curve) and without energy loss and \emph{Bohm} diffusion escape (dotted curve). We adopt parameters as follows: $\dot{N}_{0}=3.0~p^{-1}~\rm cm^{-3}~s^{-1}$, $B=0.1~\rm G$, $u=3.98~\rm erg~cm^{-3}$, $\gamma_{0}=100$, $\sigma_{\rm mag}=0.1$, $\eta=1.0$, $\xi=0.1$, $\omega=0.1$, and $r_{s}=1.0\times10^{16}~\rm cm$.}
	\label{fig:3}
\end{figure}
\section{Theoretical Photon Spectrum}
Once we have solved the steady state transport equation to determine the particle distribution in the co-moving frame of the blob in the jet, we can use the solution to calculate the jet emission components due to both synchrotron and synchrotron self-Compton emission.
\subsection{Synchrotron Emission}
Assuming an isotropic distribution of electrons, the theoretical synchrotron emission coefficient can be calculated by convolving the solution with the isotropic synchrotron emission power (e.g., Rybicki \& Lightman 1979),
\begin{eqnarray}
j_{\rm syn}(\nu)&=&\frac{\sqrt{3}e^{3}B }{4\pi m_{e}c^{2}}\int N_{G}(\gamma_{0}, \gamma)\nonumber\\&\times&R(\frac{4\pi m_{e}c\nu}{3eB\gamma^{2}})d\gamma~~\rm erg~cm^{-3}~s^{-1}~Hz^{-1}\;,
\label{Eq:47}
\end{eqnarray}
where $R(x)$ is the modified Bessel functions of 5/3 order. Synchrotron emission is accompanied by absorption, in that a photon interacts with an electron, loss its energy. According to a classical scheme of electron-dynamics, we obtain absorption
coefficient,
\begin{eqnarray}
k_{\rm syn}(\nu)&=&-\frac{\sqrt{3}e^{3}B }{8\pi m_{e}^{2}c^{2}}\int\gamma^{2}R(\frac{4\pi m_{e}c\nu}{3eB\gamma^{2}})\nonumber\\&\times& \frac{\partial}{\partial\gamma}\biggl[\frac{N(\gamma, \gamma_{0})}{\gamma^{2}}\biggr]d\gamma~~\rm cm^{-1}\;.
\label{Eq:48}
\end{eqnarray}
In the spherical geometry structure, the synchrotron intensity is given (e.g., Bloom \& Marscher 1996; Kataoka et al. 1999):
\begin{equation}
I_{\rm syn}(\nu)=\frac{j_{\rm syn}(\nu)}{k_{\rm syn}(\nu)}\biggl[1-e^{-k_{\rm syn}(\nu)r_{s}}\biggr]~~\rm erg~cm^{-2}~s^{-1}~Hz^{-1}\;.
\label{Eq:49}
\end{equation}

\subsection{Synchrotron Self-Compton Emission}
We assume a uniform synchrotron intensity in the whole radiation region, corrected for the fact that in reality it decrease along the blob radius (Gould 1979). Thus, the emission coefficient of ICs is obtained
\begin{equation}
j_{\rm ic}(\nu)=\frac{h}{4\pi}\epsilon_{\rm ic}\varrho(\epsilon_{\rm ic})~~\rm erg~cm^{-3}~s^{-1}~Hz^{-1}\;,
\label{Eq:50}
\end{equation}
where $\epsilon=h\nu/m_{e}c^{2}$ is the dimensionless particles energy, and $\varrho(\epsilon_{\rm ic})$ the differential photon production rate with
\begin{eqnarray}
\varrho(\epsilon_{\rm ic})&=&\int n(\epsilon_{\rm syn})d\epsilon_{\rm syn}\nonumber\\&\times&\int N_{G}(\gamma, \gamma_{0})\Omega(\epsilon_{\rm ic}, \gamma, \epsilon_{\rm syn})d\gamma~~\rm cm^{-3}~s^{-1}\;.
\label{Eq:51}
\end{eqnarray}
Here the number density of the synchrotron photons per energy interval, $n(\epsilon_{\rm syn})$, is described by
\begin{equation}
n(\epsilon_{\rm syn})=\frac{4\pi}{hc\epsilon_{\rm syn}}\frac{j_{syn}(\nu)}{k_{syn}(\nu)}[1-e^{-k_{syn}(\nu)r_{s}}]~~\rm cm^{-3}\;,
\label{Eq:52}
\end{equation}
and the Compton kernel $\Omega(\epsilon_{\rm ic},\gamma,\epsilon_{\rm syn})$ is given by (e.g., Jones 1968)
\begin{eqnarray}
C(\epsilon_{\rm ic},\gamma,\epsilon_{\rm syn})&=&\frac{2\pi r_{e}^{2}c}{\gamma^{2}\epsilon_{\rm syn}}\biggl[2\kappa ln\kappa+(1+2\kappa)(1-\kappa)\nonumber\\&+&\frac{(4\gamma\kappa\epsilon_{\rm syn})^{2}}{2(1+4\gamma\kappa\epsilon_{\rm syn})}(1-\kappa)\biggr]~~\rm cm^{3}~s^{-1}\;,
\label{Eq:53}
\end{eqnarray}
where $r_{e}$ is the classical electron radius, and $\kappa$ satisfies $\kappa=\epsilon_{\rm ic}/4\gamma\epsilon_{\rm syn}(\gamma-\epsilon_{\rm ic})$.

For a given $\epsilon_{\rm syn}$ and $\gamma$, differential photon production rate $\varrho(\epsilon_{\rm ic})$ can be performed under the range
\begin{equation}
\epsilon_{\rm syn}\leqslant \epsilon_{\rm ic}\leqslant \frac{4\epsilon_{\rm syn}\gamma^{2}}{1+4\epsilon_{\rm syn}\gamma}\;.
\label{Eq:54}
\end{equation}
Then, we can obtain the synchrotron self-Compton emission intensity:
\begin{equation}
I_{\rm ic}(\nu)=j_{\rm ic}(\nu)r_{s}~~\rm erg~cm^{-2}~s^{-1}~Hz^{-1}\;.
\label{Eq:55}
\end{equation}

\subsection{$\gamma\gamma$ Attenuation}
Since Very high energy (VHE) photons, general $E_{\gamma}>0.1$ TeV, from the source are attenuated by photons from the extragalactic background light (EBL), we should take the absorption effect. These scenarios give the flux density observed at the Earth as follows (e.g., Zheng \& Zhang 2011; Zheng \& Kang 2013; Zheng et al. 2018b)
\begin{equation}
F_{\rm obs.}(\nu)=\frac{\pi\delta^{3}(1+z)r_{s}^{2}}{d_{L}^{2}}\biggl[I_{\rm syn}(\nu)+I_{\rm ic}(\nu)\biggr]\times e^{-\tau(\nu,z)}\;,
\label{Eq:56}
\end{equation}
where, $d_{L}$ is the luminosity distance, $\delta$ the Doppler factor (e.g., Rybicki \& Lightman 1979), and $\tau(\nu,z)$ the absorption optical depth due to VHE photons interactions with the photons from EBL (Kneiske et al. 2004; Dwek \& Krennrich 2005).

\section{Application of Theoretical Photon Spectrum}
In this section, we apply the theoretical photon spectrum to attempt to understand the nature of the quiescent state emission from blazar jet. In order to do so, we first determine the model parameters and show the effects on theoretical photon spectrum for various parameter changes. We then apply the theoretical photon spectrum to the quiescent state emission from PKS 0414+009.
\subsection{Determination of model parameters}
Application of the theoretical photon spectrum requires the specification of both the particle spectral parameters including $\dot{N_{0}}$, $D_{0}$, $\gamma_{0}$, $\hat{A}$, $\hat{B}$, $\hat{C}$, $\hat{F}$, and the jet parameters including $B$, $\delta$, $r_{s}$. To conveniently determine the model parameters, we expect to relate the dimensionless theory parameters  $\hat{A}$, $\hat{B}$, $\hat{C}$, and $\hat{F}$ to some special physical quantity. Reminding the Eqs. (\ref{Eq:8}), (\ref{Eq:12}), (\ref{Eq:18}), (\ref{Eq:20}), and (\ref{Eq:22}), we find
\begin{equation}
\hat{A}\simeq\frac{3\eta\xi}{\sigma_{\rm mag}}\;,
\label{Eq:57}
\end{equation}
\begin{equation}
\hat{B}=5.54\times10^{-13}\frac{\eta u}{\sigma_{\rm mag}}(\frac{B}{0.1~\rm G})^{-1}\;,
\label{Eq:58}
\end{equation}
\begin{equation}
\hat{C}=\frac{3\eta}{\omega\sigma_{\rm mag}}\;,
\label{Eq:59}
\end{equation}
and
\begin{equation}
\hat{F}=8.74\times10^{-26}\eta^{2}\sigma_{\rm mag}^{-1}(\frac{r_{s}}{10^{17}~\rm cm})^{-2}(\frac{B}{0.1~\rm G})^{-2}\;.
\label{Eq:60}
\end{equation}

In our approach, we treat $\dot{N_{0}}$, $\gamma_{0}$, $\eta$, $\xi$, $\hat{A}$, $\hat{B}$, and $\hat{C}$ as free particle spectral parameters. In these scenarios, we can deduce the magnetization parameter, $\sigma_{\rm mag}$,
\begin{equation}
\sigma_{\rm mag}=\frac{3\eta\xi}{\hat{A}}\;,
\label{Eq:61}
\end{equation}
and dimensionless timescale constant, $\omega$,
\begin{equation}
\omega=\frac{\hat{A}}{\hat{C}\xi}\;.
\label{Eq:62}
\end{equation}
Once the value of $\sigma_{\rm mag}$ have been obtained, we can calculate the parameters $D_{0}$, $\hat{F}$, and both magnetic field and soft photon field energy density $u$ using Eq. (\ref{Eq:8}), Eq. (\ref{Eq:60}) and the relation
\begin{equation}
u=1.8\times10^{12}\hat{B}\sigma_{\rm mag}\eta^{-1}(\frac{B}{0.1~\rm G})~~\rm erg~cm^{-3}\;,
\label{Eq:58}
\end{equation}
by adding two jet parameters $B$ and $r_{s}$.

The model presented in this work uses a exact electron distribution that is solved from a generalized transport equation that contains the terms describing first-order and secondary-order \emph{Fermi} acceleration, escape of particle due to both the advection and spatial diffusion, energy losses due to synchrotron emission and IC scattering of an assumed soft photon field. Since it specifies the physical processes, instead of an assumed electron distribution, we have to introduce more parameters to control formation on the electron spectrum. In principle, the model requires ten free parameters ($\dot{N_{0}}$, $\gamma_{0}$, $\eta$, $\xi$, $\hat{A}$, $\hat{B}$, $\hat{C}$, $B$, $\delta$, $r_{s}$) to calculate the theoretical photon spectra. More free parameters greatly increase the uncertainty of model spectrum.

To help alleviate these problems, we establish the following constraint on the parameters: 1) Since the parameter $\eta$ is valid for the case of particles with gyroradii smaller than the correlation length of the field, this scenario imply that $\eta\leq1$; 2) the maximum efficiency factor $\xi$ is determined by ensuring that speed of shock wave $v_{s}$ is less than the light speed $c$. Hence we obtain the constraint $\xi\leq1$; 3) the size of emission region is constrained by the variability timescales $t_{\rm var}$ with $r_{s}\sim c\delta t_{\rm var}/(1+z)$.

\subsection{Effects of the changes in parameters}
In order to penetrate the variety of spectral behavious observed from blazar jets, it is important to investigate how the particle spectral parameters and/or jet parameters in the emission region effects the theoretical photon spectrum. To highlight the effects caused by the changes of individual parameters, we change only one parameter with other parameters fixed. We adopt $\dot{N_{0}}=8.0\times10^{21}~p^{-1}~\rm cm^{-3}~s^{-1}$, $\gamma_{0}=100$, $\eta=1.0$, $\xi=0.1$, $\hat{A}=30$, $\hat{B}=5.54\times10^{-8}$, $\hat{C}=66$, $B=0.1~\rm G$, $\delta=21$, and $r_{s}=5.0\times10^{15}~\rm cm$ as a baseline of the theoretical SED.

\begin{figure}
	\centering
		\includegraphics[width=9.0 cm]{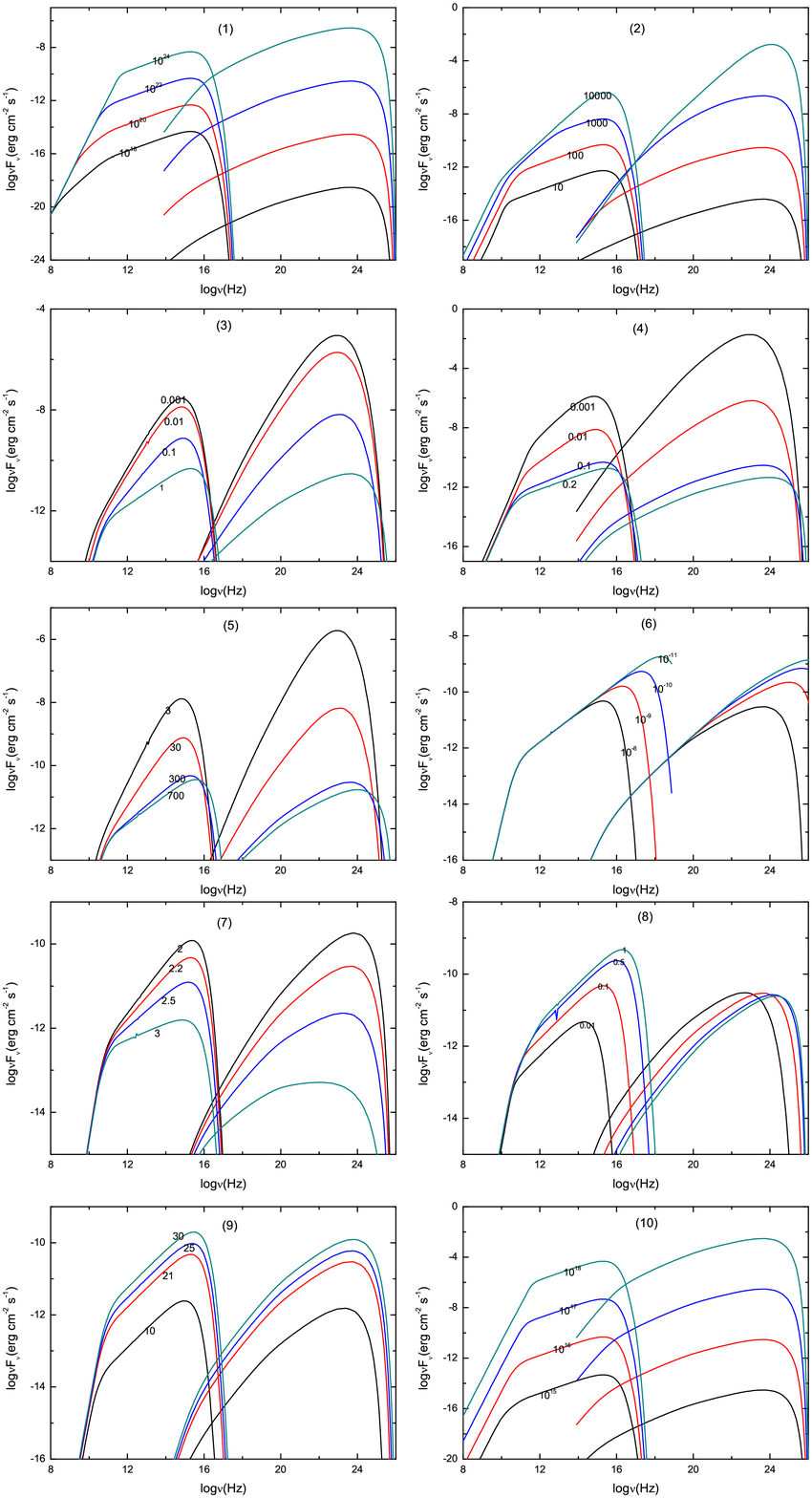}
		
	\caption{Theoretical photon spectrum for various parameter changes. (1) continual injection rate $\dot{N}$; (2) characteristic Lorentz factor of injection particle $\gamma_{0}$; (3) dimensionless parameter $\eta$; (4) shock acceleration efficiency factor $\xi$; (5) dimensionless parameter $\hat{A}$; (6) dimensionless parameter $\hat{B}$; (7) dimensionless parameter $\hat{C}$; (8) local magnetic field strength $B$; (9) beaming factor $\delta$; and (10) size of blob $r_{s}$.}
	\label{fig:4}
\end{figure}

The changes in synchrotron and SSC spectrum by varying the free parameters are shown in Figure {\ref{fig:4}}. We note that : 1) the intensity of spectrum becomes higher when $\dot{N}$ and $r_{s}$ increase, because the injected power depends on continual injection rate $\dot{N}$ and the total number of particles is proportional to the volume of the blob. The change in the flux proportional to $\delta^{4}$ and the blue shift of frequency proportional to $\delta$ are clearly seen; 2) the shape of theoretical photon spectrum are dominated by characteristic Lorentz factor of injection particle $\gamma_{0}$, dimensionless parameter $\eta$, shock acceleration efficiency factor $\xi$, dimensionless parameter $\hat{A}$, and dimensionless parameter $\hat{C}$, since these parameters determine on the particle distribution; 3) due to the peak frequency of synchrotron component proportional to magnetic field $B$, the peak frequencies of synchrotron and SSC component increase when magnetic field strengthens. On the contrary, the dimensionless parameter $\hat{B}$ increases resulting to decrease the equilibrium energy of particle, the peak frequencies of synchrotron and SSC component decrease when the dimensionless parameter $\hat{B}$ increases.

\subsection{Application to 1ES 0414+009}
1ES 0414+009 resides in an elliptical host galaxy at a redshift of $z=0.287$ (Halpern et al. 1991), with absolute magnitude $M_{R}=-23.5$ (Falomo et al. 2003). Both the original radio, optical and X-ray observations (Ulmer et al. 1983) and polarization measurements (Impey \& Tapia 1998) confirmed the classifications of this source as a BL Lac object. The archival observations of 1ES 0414+009 in X-ray bands show the synchrotron peak above a few keV (Giommi et al. 1990; Brinkmann et al. 1995; Kubo et al. 1998; Costamante et al. 2001; Sambruna et al. 2001; Beckmann et al. 2002). As an extreme source, the spectrum of BL Lac object 1ES 0414+009 was measured extended up to 0.1 TeV in the multi-wavelength observation campaign in the epoch 2005-2009 (Abramowski et al. 2012). Since the particle distribution in the context is solved from the stationary particle transport equation, what is important for the data is that the observations were made during in quiescent or averaged during in some epoches. In this scenario, we compile the archival data from Costamante \& Ghisellini (2002) and the average spectra in the epoch 2005-2009 from Abramowski et al. (2012).

\begin{figure}
	\centering
		\includegraphics[width=9.0 cm]{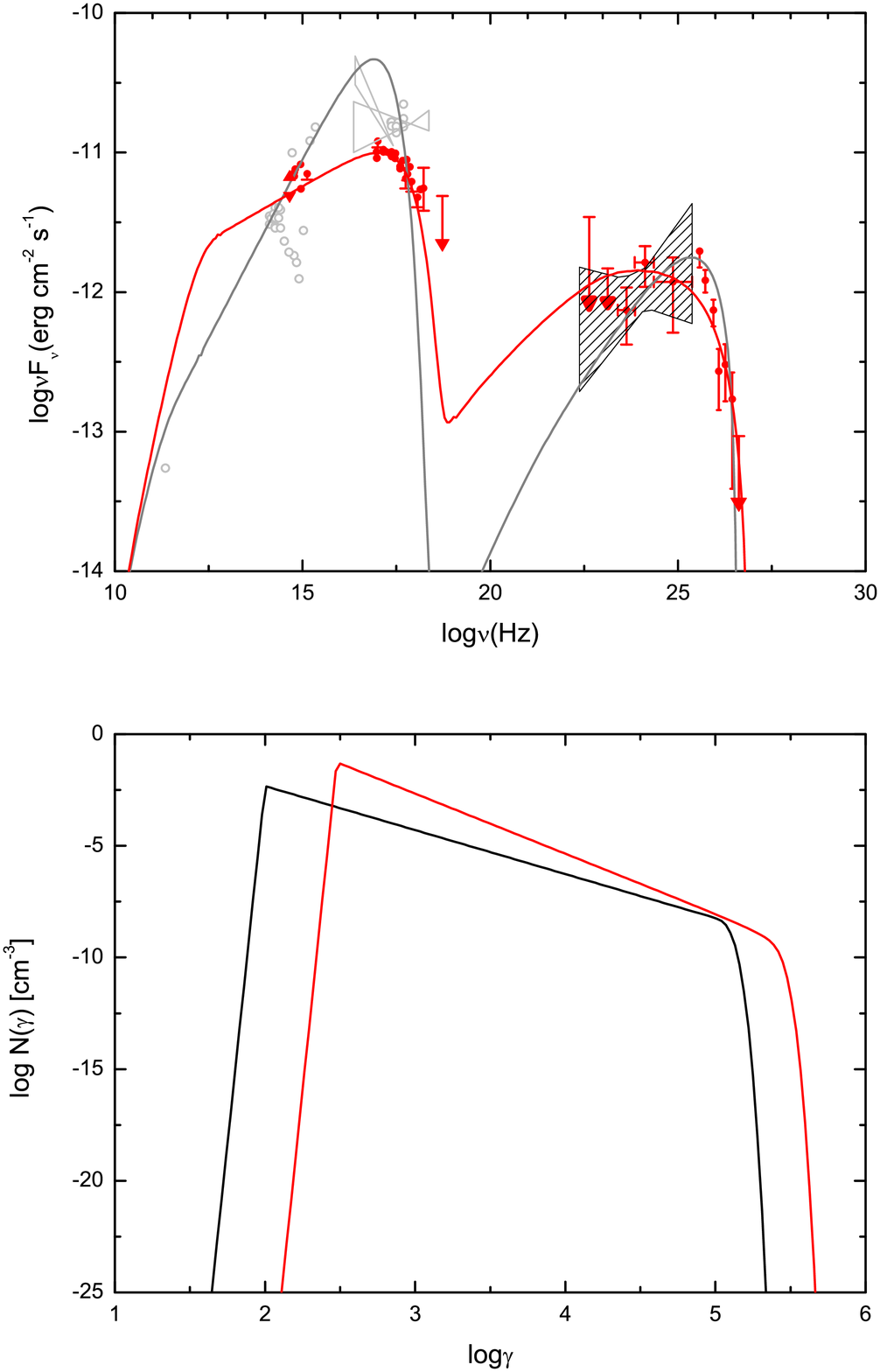}
		
	\caption{Comparisons of theoretical multi-wavelength spectra with observed data for BL Lac object 1ES 0414+009 (top panel). The plotted red and gray curves are the corresponding to theoretical multi-wavelength spectra for average data in the epoch 2005-2009 and archival data, respectively. The average SED in the epoch 2005-2009 that is taken from Abramowski et al. (2012) are shown in red. The gray points and butterflies are a collection of archival data from Costamante \& Ghisellini (2002) and references therein. The shaded area shows the \emph{Fermi} upper bounds at the 99\% confidence level. The particle distributions of reproducing the multi-wavelength spectra are shown in the bottom panel. The red and gray distributions produce the theoretical multi-wavelength spectra for average data in the epoch 2005-2009 and archival data, respectively.}
	\label{fig:5}
\end{figure}

As mentioned above, in order to check whether the scenario in the context can explain the multi-wavelength emission, we apply the results of simulation to the extreme BL Lac object 1ES 0414+009. In order to do that, we first establish the value of model parameters. Our approach for reproducing the multi-wavelength spectrum from 1ES 0414+009 sets $\eta=1$ and $\xi=0.2$ in all of the numerical calculations. The other model parameters $\dot{N_{0}}$, $\gamma_{0}$, $\hat{A}$, $\hat{B}$, $\hat{C}$, $B$, $\delta$, and $r_{s}$ are varied until a reasonable qualitative fit to the multi-wavelength spectral data is obtained. That is, we assume the the continual injection rate $\dot{N_{0}}$ with a injected Lorentz factor $\gamma_{0}$, we calculate the electron distribution with the dimensionless parameter $\hat{A}$, $\hat{B}$ and $\hat{C}$. Therefore, we can reproduce the multi-wavelength spectrum with the magnetic field strength $B$ , the Doppler factor $\delta$, and the size of emission region $r_{s}$. We report the physical parameters of both average spectra in the epoch 2005-2009 and archival data in Table 1. In Figure {\ref{fig:5}}, we compare theoretical multi-wavelength spectrum with archival data and average spectra in the epoch 2005-2009 from BL Lac object 1ES 0414+009. We also show the particle distributions of reproducing the multi-wavelength spectra. It can be seen that: 1) the analytical particle transport model considered here is able to roughly reproduce the observed spectra; 2) in despite of the model requiring on higher injection power, the particle distribution in the context is able to reproduce the multi-wavelength spectrum with reasonable assumptions about the physical parameters.

 \begin{deluxetable}{lcc}
 \tablecaption{Physical parameters of the model spectra}
 \tablewidth{0pt}
 \tablehead{
 \colhead{physical parameters} & \colhead{2005-2009 data} & \colhead{archival data}}
 \startdata
 $\dot{N_{0}}~[\rm p^{-1}~cm^{-3}~s^{-1}]$     & $1.2\times10^{21}$   & $1.0\times10^{20}$ \\
 $\gamma_{0}$                                  & 300                  & $100$              \\
 $\xi$                                         & 0.2                  & 0.2                \\
 $\eta$                                        & 1.0                  & 1.0                \\
 $\hat{A}$                                     & 60                   & 60                 \\
 $\hat{B}$                                     & $9.4\times10^{-10}$  & $4.44\times10^{-9}$\\
 $\hat{C}$                                     & 174                  & 126                \\
 $B~[\rm G]$                                   & 0.15                 & 0.15               \\
 $\delta$                                      & 31                   & 31                 \\
 $r_{s}~[\rm cm]$                              & $1.7\times10^{16}$   & $3.8\times10^{16}$ \\
 $\hat{F}$                                     & $1.34\times10^{-22}$ & $2.69\times10^{-23}$  \\
 $D_{0}~[\rm s^{-1}]$                          & $8.79\times10^{3}$   & $8.79\times10^{3}$  \\
 $\sigma_{\rm mag}$                            & 0.01                 & 0.01  \\
 $\omega$                                      & 1.72                 & 2.38  \\
 $u~[\rm erg~cm^{-3}]$                         & 25.38                & 119.88  \\
 $P_{\rm inj}~[\rm erg~s^{-1}]$                & $1.6\times10^{50}$   & $5.14\times10^{49}$  \\
\enddata
 \end{deluxetable}

\section{Discussion}
As an open issue, determining the jet physics from the SED is a tricky problem of inversion. The present paper introduce a analytical particle transport model to reproduce quiescent broadband SED of blazar. In the model, the exact electron distribution is solved from a generalized transport equation that contains the terms describing first-order and secondary-order \emph{Fermi} acceleration, escape of particle due to both the advection and spatial diffusion, energy losses due to synchrotron emission and IC scattering of an assumed soft photon field. We don't take into account modification the electron distribution in the Klein - Nishina (KN) regime (e.g. Moderski et al. 2005; Nakar et al. 2009). Furthermore, we don't include the non-linear synchrotron (Schlickeiser \& Lerche 2007) and SSC (Schlickeiser 2009) cooling of relativistic electrons. Assuming suitable model parameters, we apply the results of simulation to the extreme BL Lac object 1ES 0414+009. It is clear that the particle injection rata, $\dot{N}$, and the Lorentz factor of injected electrons, $\gamma_{0}$, play an important role in determining an emission intensity. Assuming isotropic emission, the associated power in the injected particles is given by $P_{\rm inj}=4\pi r_{\rm s}^{3}\gamma_{0}m_{e}^{2}c^{3}\dot{N}/3$. The model presented in this work suggests a extreme injection power with $P_{\rm inj}\sim 10^{50}~\rm erg~s^{-1}$. This value exceeds the Eddington luminosity with a supper-massive black hole $\sim2\times10^{9}M_{\odot}$ in two orders of magnitude. In despite of the equilibrium between the radiation pressure acting outward and the gravitational force acting inward is ruled for a spherically symmetric geometry (Eddington 1916), if the photon are trapped inside the accretion flows and are advected into the black holes (e.g., Abramowicz et al. 1988; Beloborodov 1998; Wang et al. 1999; Mineshige et al. 2000; Chen \& Wang  2004), above which the radiation force dominates the gravity of the central black hole, the radiation luminosity can exceed the Eddington luminosity. In this paper, we do not propose an explanation for why the injection power exceeds the Eddington luminosity. However, it is interesting to speculate that this might be a result of the shock front over-runs a region in the jet in which the local plasma density is enhanced (e.g., Kirk et al. 1998; Zheng \& Zhang 2011). In this scenario, we expect to that the number of injection particles increase as an avalanche occurs in the jet. Incidentally, the particle injection rate, $\dot{N}$, increases significantly, and results in a extreme injection power into the emission region. We also noted that the model suggests the magnetization parameter $\sigma_{\rm mag}=0.01$. This result is within the range from $\sigma_{\rm mag}\sim0.001$ in the MHD models (e.g., Kennel \& Coroniti 1984) to $\sigma_{\rm mag}\sim1$ in the striped wind models (e.g., Komissarov 2003).

The present work differs from the earlier efforts that assume only that the escape of the particles occurs via spatial diffusion (e.g., Stawarz \& Petrosian 2008; Tammi \& Duffy 2009; Lewis et al. 2016; 2018). In the context, we concentrate on both the shock regulated escape and \emph{Bohm} diffusive escape. We suggest that the advection is a significant escape mechanism in blazar jet. Since the advection tends to harden the particle distribution, which enhances the high energy components of resulting synchrotron and SSC spectrum from blazar jet, we argue that the model can likely be used to comprehend on the origin of hard spectra, while there are some other interpretations on observed hard spectra from distant blazars (Lefa et al. 2011; Zheng \& Kang 2013; Cerruti et al. 2015; Zheng et al. 2016).

The analytical particle transport model is based on the development of exact analytical solutions to the linear transport equation. A potential drawback of the model is that the SSC losses can not include into establishing the particle distribution, since they are inherently nonlinear. To render the nonlinear effect of SSC process, we assume a constant soft photon field to instead of the synchrotron emission field. On the bias of the model results, we can also calculate the synchrotron emission field $u_{syn}=3.45\times10^{-4}~\rm erg~cm^{-3}$ for 2005-2009 data and $u_{syn}=1.64\times10^{-4}~\rm erg~cm^{-3}$ for archival data, respectively. If we directly include the synchrotron emission field into the transport equation, we find $u=u_{B}+u_{syn}\sim10^{-3}~\rm erg~cm^{-3}$. This value is less four orders of magnitude than the constant soft photon fields of the model assumed. It is believed that the electron populations where the energy is around the equilibrium energy produce the X-rays spectra. The observation shows the SED of source with a synchrotron peak energy locating at 0.1 keV (Abramowski et al. 2012). In this scenario, we can consider that the synchrotron emission of the electron populations with equilibrium energy contribute on the most of intensity around the synchrotron peak. These issues imply that the equilibrium Lorentz factor satisfies $\gamma_{e}\sim(\nu_{syn,p}/3.7\times10^{6}B\delta)^{1/2}\sim10^{5}$. A lower soft photon field results to larger equilibrium Lorentz factor. The calculated synchrotron emission field induces the equilibrium Lorentz factor $\gamma_{e}$ around $10^{7}\sim10^{8}$. This is far from the equilibrium Lorentz factor that are required by the observed synchrotron peak energy. We argue that, in despite of the assumed IC losses can not sufficiently approximate the condition, it makes sure the analytical particle transport model can obtain suitable equilibrium Lorentz factor by a kind of competition between cooling and acceleration in the case of adopting a reasonable magnetic field parameter. Leaving out of the particle escape, we expect a large soft photon field to generate the suitable equilibrium.

Actually, electrostatic acceleration is of some interest, because it characterizes the strength by which magnetic reconnection provides a source of acceleration to the leptons. As proposed in Kroon et al. (2016), electrostatic acceleration in an electric field of strength $E$, generated in the magnetic reconnection region around the shock, results in a constant momentum gain rate given by  $\dot{p}_{\rm elec}=eE$. Combining the momentum gain rate by shock acceleration Eq. (11), we can establish the first-order momentum gain rate, $\dot{p}_{\rm gain}=\dot{p}_{\rm elec}+\dot{p}_{\rm sh}=A_{0}m_{e}c$ with the first-order momentum gain rate $A_{0}=A_{\rm sh}+A_{\rm elec}$ in the unit of $\rm s^{-1}$, appearing in the Eq. (2). Where $A_{\rm elec}=eE/(m_{e}c)=1.76\times10^{6}(E/B)(B/0.1~\rm G)~\rm s^{-1}$. Reminding the process of resulting to the dimensionless model parameter $\hat{A}$, we yield the relation, $E/B=\hat{A}\sigma_{\rm mag}/(3\eta)-\xi$. It is convenient to find the relative contribution from shock acceleration and electrostatic acceleration in the emission region by the relation, $\dot{p}_{\rm elec}/\dot{p}_{\rm sh}=\xi^{-1}E/B$. It can be seen that, since the model simulations suggest $\xi<1$, if it satisfies $E/B>1$, the electrostatic acceleration dominates on the first-order momentum gain in the region. The efficient electrostatic acceleration can result to a larger equilibrium Lorentz factor and a harder particle spectrum. Conversely, if the ambient magnetic field exceeds the electric field, the shock acceleration dominates on the first-order momentum gain in the region. The model presented in this work requires increasing the parameter $u_{\rm ph}$ as free parameter to calculate the value of $E/B$. In order to simplify the model, we tempt to issue a study using the model where one not take into account the electrostatic acceleration.

It is noted that in the regime of the Lorentz factor far below the equilibrium Lorentz factor, the particle distribution is well represented by a broken power law. A particular interest is the case of without the random motions of the MHD waves, the equation is left with only the contribution due to shock acceleration. This scenario corresponds to the limit of momentum diffusion rate constant $D_{0}\rightarrow0$. After some algebra, we convenient to deduce the power-law index of the electron distribution for the case of shock acceleration as $\alpha_{\rm sh}=\lim \limits_{D_{0}\rightarrow0}\alpha_{2}=-\hat{C}/\hat{A}$ (e.g., Kroon et al. 2016). The particle in cell simulations in the regions of magnetic reconnection near the termination shock give the high energy power-law index $\alpha_{\rm sh}$ in the range of $\alpha_{\rm sh}\in[-3, -2]$ (Cerutti et al. 2014). Our model fits suggest that, $\hat{C}/\hat{A}=2.9$ and 2.1 in the epoch 2005-2009 and archival data, respectively, providing an important shock acceleration diagnostic.

Reminding the condition of resulting to the steady-state Green's function, we endeavor to propose the following paradigm for the broken power law in blazar region (e.g, Zheng et al. 2018b): in the case of a non-relativistic and parallel shock, we assume the particle distributions satisfy $N_{\rm 1}(\gamma)\propto\delta(\gamma-\gamma_{0})$ in the upstream region, where the $\gamma_{0}$ is characteristic energy of the particles. If the size of shocked flow is limitless, using the zero flux boundary condition, we could obtain a steady particle spectrum in the downstream flow $N_{\rm 2}(\gamma)=N_{0}\gamma^{-s}$ with the spectra index $s$ (e.g., Dermer \& Menon 2009). As for above scenarios result to more hard spectra than the distribution with energy loss and \emph{Bohm} diffusion escape, we leave the application and discussion in details in the future work.



\acknowledgments
\section*{acknowledgments}
We thank the anonymous referee for valuable comments and suggestions. This work was partially supported by the National Natural Science Foundation of China (Grant Nos. 11673060, 11763005 and 11873043), the Key Research Program of the Chinese Academy of Sciences (grant No. KJZD-EW-M06), the Strategic Priority Research Program ``The emergence of Cosmological Structure'' of the Chinese Academy of Sciences (grant No. XDB09000000), and the Natural Science Foundation of Yunnan Province (Grant Nos. 2016FB003 and 2017FD072). Additional support was provided by the Specialized Research Fund for Shandong Provincial Key Laboratory (Grant No. KLWH201804), by the Key Laboratory of Particle Astrophysics of Yunnan Province (Grant No. 2015DG035) and by the Research Foundation for Scientific Elitists of the Department of Education of Guizhou Province (Grant No. QJHKYZ[2018]068).

\clearpage

\end{document}